\documentclass[aps,prc,twocolumn,showpacs,showkeys,preprintnumbers,amsmath,amssymb,a4paper]{revtex4}
\usepackage{graphicx}
\usepackage{bm}
\usepackage{epsfig}

\newcommand{\be}{\begin{eqnarray}}
\newcommand{\ee}{\end{eqnarray}}

\newcommand{\dd}{\mbox{\rm d}}
\def\bfg #1{{\mbox{\boldmath $#1$}}}
\newcommand{\fmn}[2]{\mbox{${\textstyle \frac{#1}{#2}}$}}

\begin{document}
\title{Dynamics of $^{1}S_0$ diproton formation in the
$pd\to \{pp\}_sn$ and  $pN\to \{pp\}_s\pi$ reactions in the GeV
region}

\author{Yu.N.~Uzikov$^1$, J.~Haidenbauer$^2$, C.~Wilkin$^3$}

\affiliation{
$^1$Laboratory of Nuclear Problems, Joint Institute for Nuclear
Research, 141980 Dubna, Russia\\
$^2$Institut f\"ur Kernphysik (Theorie), Forschungszentrum J\"ulich,
52425 J\"ulich, Germany\\
$^3$Physics and Astronomy Department, UCL, London, WC1E 6BT, UK
}
\date{\today}

\begin{abstract}
Mechanisms for the production of $^{1\!}S_0$ diproton pairs,
$\{pp\}_{\!s}$, in the $pd\to \{pp\}_{\!s}n$ reaction are studied
at proton beam energies 0.5 -- 2\,GeV in kinematics similar to
those of backward elastic $pd$ scattering. This reaction provides
valuable information on the short-range $NN$ and $pd$ interactions
that is complementary to that investigated in the well known
$pd\to dp$ and $dp\to p(0^\circ)X$ processes. The $pd\to
\{pp\}_{\!s}n$ reaction is related to the subprocesses $\pi^0 d\to
pn$ and $pN\to \{pp\}_{\!s} \pi$ using  two different
one--pion--exchange (OPE) diagrams. Within both these models a
reasonable agreement could be obtained with the data below 1\,GeV.
The similar energy dependence of the $pd\to\{pp\}_{\!s}n$ and
$pd\to dp$ cross sections and the small ratio of about 1.5\% in
the production of $\{pp\}_{\!s}$ to deuteron final states follow
naturally within the OPE models.
\end{abstract}


\maketitle

\section{Introduction}
\begin{figure}[hbt]
\mbox{\epsfig{figure=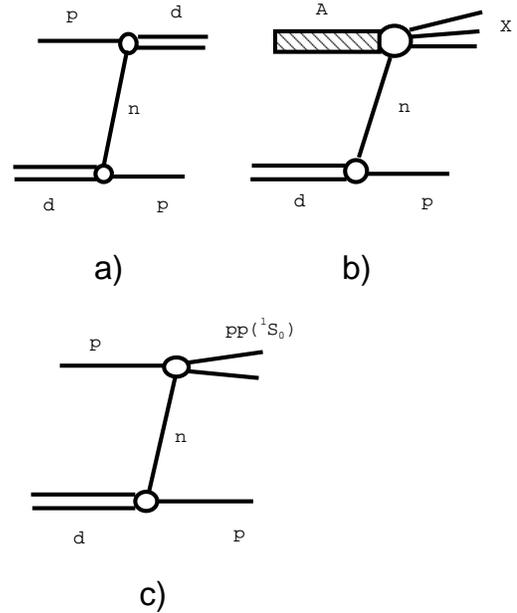,height=0.35\textheight, clip=}}
\caption{The one nucleon exchange (ONE)  mechanisms of the
reactions (a)~$pd \to  dp$, (b)~$dA\to p(0^{\circ})X$, and (c)~$pd
\to \{pp\}_{\!s}n$.} \label{figone}
\end{figure}

There is a long standing problem connected with understanding the
mechanism of proton--deuteron backward elastic scattering at
energies above 0.5\,GeV. This can be formulated as follows. Except
in the $\Delta$--isobar region of $0.4-0.6$\,GeV, the unpolarized
differential cross section $\dd\sigma/\dd\Omega(pd\to
dp)_{\theta_{cm}=180^\circ}$ can be explained qualitatively within
the impulse approximation (IA) up to large nucleon momenta in the
deuteron $k\approx1$\,GeV/c, whereas the experimental values of
the tensor analyzing power $T_{20}$ are in strong contradiction to
the IA calculations already for
$k>0.3$~GeV/c~\cite{Arvieuxpddp,punjabipddp,azhgireypddp}. Here IA
means the one--nucleon--exchange (ONE) mechanism of
Fig.~\ref{figone}a which, if it dominated the unpolarized cross
section, would allow one to probe directly the high--momentum
components in the deuteron wave functions.

A very similar problem arises in the analysis of the inclusive
disintegration of the deuteron on nuclear targets, $dA\to
p(0^\circ)X$, when the ONE mechanism of Fig.~\ref{figone}b is used
to describe the
process~\cite{perdrisatbreak,ableevbreak,azhgireybreak}. In
contrast, the tensor polarization $t_{20}$ of the recoil deuteron
in elastic electron--deuteron scattering follows very well the IA
predictions~\cite{gilmangross} up to very high transferred momenta
$Q=1.3$~GeV/c, \emph{i.e.}\ up to $k\sim Q/2=0.65$~GeV/c, if
realistic phenomenological $NN$
potentials~\cite{nijmegen,argonne,cdbonn} are used to describe the
deuteron. Corrections from
meson--exchange currents are sizable, but do not change the
picture qualitatively~\cite{arenhovel}. We must conclude that in
exclusive and inclusive $pd$ collisions at high transferred
momenta we are dealing, not only with the short-range structure of
the deuteron, but also with the specific dynamics of the $pd$
interaction and that these dynamics are entirely different from
those in the $ed\to ed$ process.

The above contradictions, referred to as the $T_{20}$-puzzle, can
be ascribed, in part, to contributions from the excitation of
nucleon isobars ($\Delta, N^*$) in the intermediate state, which
were neglected within the IA
analysis~\cite{Arvieuxpddp,punjabipddp,azhgireypddp,perdrisatbreak,ableevbreak,azhgireybreak}.
For example, the $\Delta$--mechanism seems to dominate the large
angle unpolarized $pd\to dp$ cross section in the 0.4--0.6~GeV
interval~\cite{lak,boudardillig,uzikovrev}. However, the spin
structure of the three--body forces related to the $\Delta$-isobar
is far from well established~\cite{sakai}. This therefore leads to
ambiguities in any explanation of $T_{20}$ when the
$\Delta$--isobar is included in the transition
amplitude~\cite{lak,boudardillig,uzikovrev}. It was suggested
that, in order to clarify the role of the $\Delta-$isobar, the
$pd\to \{pp\}_{\!s}n$ reaction should be
studied~\cite{imuz90,smiruz,uzjpg}. Due to isospin invariance, the
$\Delta$-mechanism is diminished by a factor of nine in the
$pd\to\{pp\}_{\!s}n$ cross section as compared to that of $pd-dp$,
whereas the ONE mechanism does not suffer a similar
suppression~\cite{uzjetf}. Therefore, the comparison of the two
reactions might allow one to get a clearer picture of the relative
importance of the ONE and $\Delta$-contributions.

The unpolarized $pd\to \{pp\}_{\!s}n$ differential cross section
has been measured for large neutron c.m.\ angles with respect to
an incident proton beam which had laboratory kinetic energies in
the range $0.6-1.9$\,GeV~\cite{komarov2003}. The predominance of
the $^{1\!}S_0$ state was guaranteed by selecting diproton events
with excitation energy $E_{pp}<3$\,MeV. An analysis of these data
was performed within a model, originally suggested to describe the
$pd\to dp$ reaction~\cite{lak}, that included one--nucleon
exchange (ONE) (Fig.~\ref{figone}c), single $pN$ scattering, and
double scattering with the excitation of the
$\Delta$--isobar~\cite{haidenuz2003}. This showed that the
contribution of the ONE mechanism in Born approximation is
actually quite small for a wide range of commonly used $NN$
potentials. Only for a soft $NN$ potential, such as the CD
Bonn~\cite{cdbonn}, and with absorptions taken into account in the
initial and final states, can a qualitative agreement with data be
achieved~\cite{haidenuz2003}. In the other extreme, harder
$NN$-potentials, \emph{e.g.}\ the Paris~\cite{paris} or especially
the Reid soft core~\cite{reid}, generate intense high--momentum
components in the $NN$ wave functions and therefore lead to very
large ONE contributions that are in strong disagreement with the
$pd\to \{pp\}_{\!s}n$ data~\cite{komarov2003}. This is the most
interesting observation resulting from the $pd\to \{pp\}_{\!s}n$
analysis of Ref.~\cite{haidenuz2003}.

\begin{figure}[hbt]
\mbox{\epsfig{figure=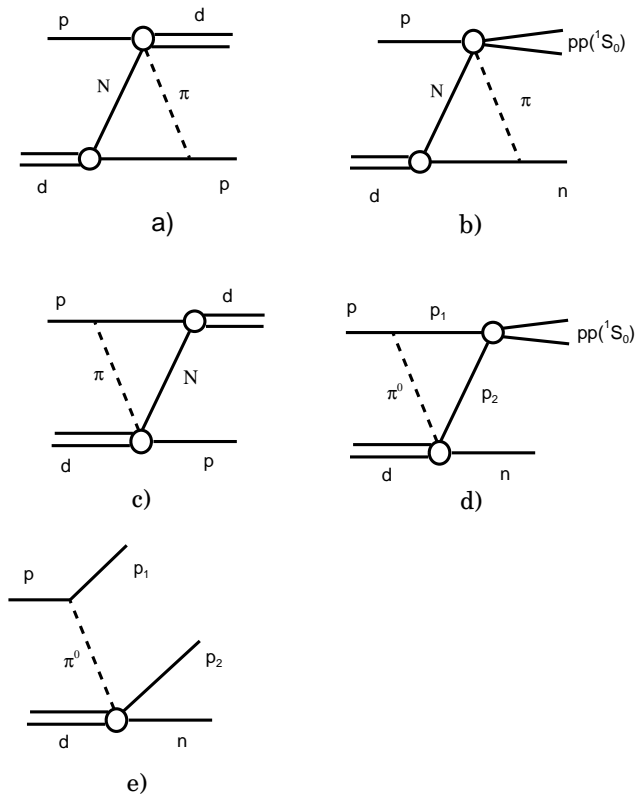,height=0.45\textheight, clip=}}
\caption{ The one--pion--exchange (OPE) mechanisms considered for
the reactions $pd \to dp$ (a,c) and  $pd \to  \{pp\}_{\!s}n$
(b,d,e): OPE-I -- a,b; OPE-II -- c,d,e. } \label{figope1}
\end{figure}

On the experimental side, the next step towards unraveling the
dynamics of the $pd\to \{pp\}_{\!s}n$ reaction will be the
measurement of the deuteron tensor analyzing power
$T_{20}$~\cite{SPINCOSY}. On the theoretical front, an important
task is to study other mechanisms that are less sensitive to high
$NN$ momentum components than the ONE mechanism. A new and
independent analysis of the $pd\to\{pp\}_{\!s}n$ dynamics has been
made possible through the recent publication of data on the $pp\to
\{pp\}_{\!s}\pi^0$ reaction~\cite{dymov06}.

In this paper we analyze the mechanisms of the $pd\to
\{pp\}_{\!s}n$ reaction that are connected with two--step
processes involving the creation and absorption of pions in the
intermediate state. The one--pion--exchange (OPE) triangle diagram
depicted in Fig.~\ref{figope1}a, and here denoted as OPE-I, was
initially invoked to describe the large angle $pd\to dp$
reaction~\cite{cwilkin69}. Here the $pd\to dp$ cross section was
connected to that for the $pp\to d\pi^+$ sub process at the same
beam energy. The predictions of the model were found to be in
qualitative agreement with the data on the energy dependence of
the $pd$ backward elastic scattering around 0.5--1.0\,GeV. An
important role of the OPE mechanism, through the
$p\{NN\}\to\,^{3\!}\textrm{He}\,\pi$ sub processes, was also found
in the reaction $p\,^{3\!}\textrm{He}\to\,^{3\!}\textrm{He}p$ at
0.5--1\,GeV~\cite{uzhaidprc}. To apply the analogous mechanism of
Fig.~\ref{figope1}b to the $pd\to \{pp\}_{\!s}n$ reaction we need
to know the amplitudes for both $pp\to\{pp\}_{\!s}\pi^0$ and
$pn\to \{pp\}_{\!s}\pi^-$. At present, however, only the
unpolarized cross section for $pp\to\{pp\}_{\!s}\pi^0$ was
measured at 0.8\,GeV~\cite{dymov06}. In the absence of data on
$\pi^-$ production, we have to make assumptions about the
$pN\to\{pp\}_{\!s}\pi$ mechanism in order to add coherently the
contributions from the $pp\to \{pp\}_{\!s}\pi^0$ and
$pn\to\{pp\}_{\!s}\pi^-$ sub processes.  The mechanisms used in the
present analysis are depicted in Fig.~\ref{fig3mech}. We show in
Sec. II that the results of the calculation within OPE-I depend
strongly on the mechanism assumed.

Such an ambiguity does not, however, appear for mechanisms with
the $\pi^0 d\to  pn$ sub process (Fig.~\ref{figope1}d and e),
which we refer to as OPE-II and discuss in Sec. III. Due to
time-reversal invariance, the predictions of OPE-I and OPE-II
would be the same for the unpolarized $pd\to dp$ cross section,
though this identity does not extend to the analyzing powers.
However, to avoid double--counting, one should never consider
together the diagrams in Fig.~\ref{figope1}a and c, since they may
be but different approximations to the same underlying physics. We
finally consider in Sec. IV the role of baryon (or Reggeon) exchange
in these reactions,  that is motivated in part by the results of
the recent measurement of the $pp\to \{pp\}_{\!s}\pi^0$
reaction~\cite{dymov06}. Numerical results for the different
models and the comparison with experiment are presented in Sec. V and
our conclusions in Sec. VI.

\section{The OPE-I mechanism}
\setcounter{equation}{0}

\begin{figure}[hbt]
\mbox{\epsfig{figure=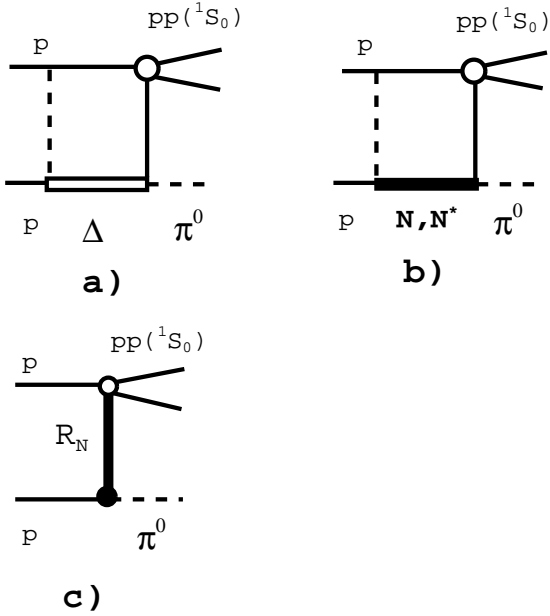,height=0.35\textheight, clip=}}
\caption{ Possible mechanisms for the $pp\to \{pp\}_{\!s}\pi^0$
reaction: (a) $\Delta$-isobar excitation in $\pi N$-rescattering,
(b) $\pi N$- rescattering in the $T=\frac{1}{2}$ state (nucleon or
$N^*$ in the $s$--channel), (c) $T=\fmn{1}{2}$ baryon ($N,\, N^*$)
or Reggeon exchange (BRE) in the $t$--channel.} \label{fig3mech}
\end{figure}

In the OPE-I approach to the $pd\to\{pp\}_{\!s}n$ reaction, the
sub process $pN\to\{pp\}_s\pi$ is invoked but, as shown in
Fig.~\ref{figope1}b, there are contributions with either a $\pi^0$
$(A^{0})$ or a $\pi^-$ meson ($A^{-}$) in the intermediate
state. The coherent sum of these diagrams depends on the
contribution of $T=1/2$ exchange in the pion--production
amplitude. Using the mechanisms depicted in Fig.~\ref{fig3mech},
and assuming isospin invariance, we obtain the following results
for the deuteron breakup amplitude:


\begin{widetext}
\begin{equation}
\label{trimech} A^{0}({pd\to \{pp\}_{\!s}n})+A^{-}({pd\to
\{pp\}_{\!s}n})=
\begin{cases}
    2A^{0}({pd\to \{pp\}_{\!s}n}),
 \ {\text{ $\Delta$ in $\pi N$-rescattering, Fig.~\ref{fig3mech}a }} \\
 -A^{0}({pd\to \{pp\}_{\!s}n}),
 \ {\text { $N$ or $N^*$ in $\pi N$ rescattering, Fig.~\ref{fig3mech}b}}\\
 3A^{0}({pd\to \{pp\}_{\!s}n}),
 \ {\text {$T=\fmn{1}{2}$ baryon exchange in $t$--channel,
 Fig.~\ref{fig3mech}c }}\\
\end{cases}
\end{equation}
\end{widetext}

The evaluation of the $A({pd\to \{pp\}_{\!s}n})$ amplitude of
Fig.~\ref{figope1}b can be performed using a similar treatment to
that of Ref.~\cite{kolybsm} for the OPE diagram in $pd\to dp$
(Fig.~\ref{figope1}a). The resulting c.m.\ differential cross
section has the form

\begin{eqnarray}
\lefteqn{\frac{\dd\sigma}{\dd\Omega}^{OPE-I}(pd\to
\{pp\}_{\!s}n)=}\nonumber \\%
\nonumber &&C_j \frac{p_f}{p_i}\frac{q_{pp}}{q_{\pi
\{pp\}}}\frac{s_{pp}}{s_{pd}} \frac{f^2_{\pi
NN}}{m^2_{\pi}}\frac{E_n+m}{E^2_n}4m^2F_{\pi}^2(k^2_\pi)\\
&&\times \Bigl\{|Z_0|^2+|Z_2|^2\Bigr \}
\frac{\dd\sigma}{\dd\Omega}(pp\to \{pp\}_{\!s}\pi^0),
\label{sope2}
\end{eqnarray}
where $f_{\pi NN}$ is the $\pi NN$ coupling constant, with $f_{\pi
NN}^2/4\pi=0.0796$, $m_{\pi}$ and $m$ are the masses of the pion
and the nucleon, respectively, $k_{\pi}$ is the four--momentum of
the virtual pion,
$F_{\pi{}NN}(k_{\pi}^2)=(\Lambda^2-m_{\pi}^2)/(\Lambda^2-k^2_{\pi})$
is the $\pi NN$ form factor, $E_n$ is the total energy of the
final neutron in the laboratory system, $p_i$ and $p_f$ are c.m.\
momenta in the initial and final states of the reaction $pd\to
\{pp\}_{\!s}n$, respectively, $s_{ij}$ is the squared invariant
mass, and $q_{ij}$ is the relative momentum in the system $j+i$.
It is assumed that the cross sections on the left and right hand
sides of Eq.~(\ref{sope2}) are to be taken at the same beam energy
and the c.m.\ production angle of the neutron and $\pi^0$ are both
equal to $180^\circ$. The beam energy for the reaction $pp\to
\{pp\}_s\pi$ determines uniquely the values $s_{pp}$, $q_{pp}$ and
$q_{\pi\{pp\}}$.

The coefficient $C_j$ ($j=a,b,c$) depends on the isospin
dependence of the $NN\to \{pp\}_{\!s}\pi $ reaction. Using
Eq.~(\ref{trimech}) we find for the mechanisms depicted in
Fig.~\ref{fig3mech}a, b and c, respectively, $C_a=1$, $C_b= 4$ and
$C_c=9$. The transition form factors $Z_L$ are defined through
\begin{eqnarray}
\label{Z02} \lefteqn{Z_0=\kappa |\mathbf{p}_n|F_0(p_I)-i\Phi_{10}(p_I,\delta_I),}\\%
&&\hspace{-6mm}Z_2= \kappa
|\mathbf{p}_n|F_2(p_I)-\frac{i}{\sqrt{5}} \Bigl
[\sqrt{3}\Phi_{32}(p_I,\delta_I)
-\sqrt{2}\Phi_{12}(p_I,\delta_I)\Bigr ],\nonumber
\end{eqnarray}
where
\begin{eqnarray}
\nonumber
F_L(p_I)&=&i^L\int_0^\infty j_L(p_Ir)u_L(r)\exp{(-\delta_I r)}\,r\,\dd r,\\
\Phi_{lL}(p_I,\delta_I)&=& i^l\int_0^\infty
j_l(p_Ir)u_L(r)(1+\delta_I r)\exp{(-\delta_I r )}\,\dd r,
\nonumber \\ \label{phi02}
\end{eqnarray}
and $u_0(r)$ and $u_2(r)$ are the S-- and D--state components of
the deuteron wave function, respectively, normalized as
\begin{eqnarray}
\label{normar} 4\pi\,\int_0^\infty[u_0^2(r)+u_2^2(r)]\,r^2\dd r=1.
\end{eqnarray}
In Eqs.~(\ref{phi02}), $j_l(pr)$ is the spherical Bessel function.
Kinematical variables are defined as
\begin{eqnarray}
\delta^2_I=\frac{T_n^2}{(E_n/m)^2}+\frac{m^2_{\pi}}{E_n/m},\ \
\label{deltaiII} \nonumber \\
\label{kinem2} \kappa=-\frac{m}{E_n}\frac{T_n}{E_n+m},
\mathbf{p}_I=\frac{\mathbf{p}_n}{E_n/m},
\end{eqnarray}
where $E_n$, $\mathbf{p}_n$ and $T_n=E_n-m$ are the total energy,
three--momentum, and kinetic energy of the final neutron in the
rest frame of the initial deuteron.

For the $pd\to dp$ reaction, the sum of the OPE-I amplitudes with
the $\pi^0$ and $\pi^+$ mesons in the intermediate state is
$A^{\pi^0}_{pd\to dp}+A^{\pi^+}_{pd\to dp}= 3A^{\pi^0}_{pd\to
dp}$, independent of the model for pion production, as found also
in Ref.~\cite{vegh}. Using this result with Eq.~(\ref{trimech}),
and neglecting the difference between the masses of the deuteron
and diproton, there is a relation between the c.m.\ cross sections
of the $pd\to \{pp\}_{\!s}n$ and $pd\to dp$ reactions within the
OPE-I model:
\begin{equation}
\label{relation2} \frac{\dd\sigma}{\dd\Omega}^{OPE-I}(pd\to
\{pp\}_{\!s}n)= R_I\times
\frac{\dd\sigma}{\dd\Omega}^{OPE-I}(pd\to dp).
\end{equation}
The factor $R_I$ depends on the mechanism of pion production
depicted in Fig.~\ref{fig3mech} through
\begin{equation}
\label{ratio1}
R_I=
\begin{cases}
    \frac{4}{9}r,
 \ {\text{Fig.~\ref{fig3mech}a,}} \\
  \frac{1}{9}r,
 \ {\text {Fig.~\ref{fig3mech}b,}}\\
 \phantom{w}r,
 \ {\text {Fig.~\ref{fig3mech}c,}}\\
\end{cases}
\end{equation}
where $r$ is the ratio
\begin{equation}
\label{rsmall} r=\left.\frac{\dd\sigma}{\dd\Omega}(pp\to
\{pp\}_{\!s}\pi^0)\right/\frac{\dd\sigma}{\dd\Omega}(pn\to
d\pi^0).
\end{equation}
The cross sections in Eq.~(\ref{rsmall}) are to be taken at the
same beam energy and scattering angle.

\section{The OPE-II mechanism}
\setcounter{equation}{0}

In the OPE-II approach, the deuteron breakup is driven by the $\pi
d\to pN$ sub process. The contribution of the diagram of
Fig.~\ref{figope1}c to $pd$ backward elastic scattering, as well
as to the $pd\to \{pp\}_{\!s}n$ reaction, were not considered in
Refs.~\cite{cwilkin69,kolybsm,vegh}. We therefore analyze these
amplitudes in somewhat greater detail.

\subsection{The $pd\to \{pp\}_{\!s}n$ reaction}

For the deuteron breakup reaction $pd\to \{pp\}_{\!s}n$, we
consider the sum of the two diagrams shown in Fig.~\ref{figope1}d
and \ref{figope1}e. The $pp\pi^0$ vertex function is
\begin{eqnarray}
A_{\nu_p}^{\nu_1}(p\to \pi^0 p_1)=\frac{f_{\pi NN}}{m_{\pi}}
<\chi_{\nu_1}|{\bfg \sigma}\cdot \mathbf{Q}|\chi_{\nu_p}>\nonumber\\
\label{pppi0} \times({\bfg \tau}\cdot {\bfg \phi}_{\pi}) 2mF_{\pi
NN}(k_{\pi}^2).
\end{eqnarray}
Here $\bfg \sigma$ and  $\bfg \tau$ are the Pauli matrices for
spin and isospin, respectively, $\chi_{\nu_i}$ is the Pauli spinor
with $\nu_i$ being the $z$-projection of the spin of the
\emph{i}'th proton ($i=1,p$), ${\bfg \phi}_{\pi}$ is the isospin
state of the pion, and $\mathbf{Q}$ is the three--momentum defined
as
\begin{eqnarray}
 \mathbf{Q}=\sqrt{\frac{E_p+m}{E_{p_1}+m}}\mathbf{p}_1
- \sqrt{\frac{E_{p_1}+m}{E_{p}+m}}\mathbf{p}_p\nonumber \\ \label{momQ}
 \approx
\sqrt{\frac{E_p+m}{E_{p_1}+m}}(\mathbf{p}_1-\frac{2m}{E_p+m}\mathbf{p}_p),
\end{eqnarray}
with $\mathbf{p}_i$ and $E_i$ being the momentum and total energy
of the \emph{i}'th proton. The half--off--shell $pp$ scattering
amplitude is (see, for example, Ref.~\cite{smiruz})
\begin{eqnarray}
\nonumber \lefteqn{A_{\nu_1 \nu_2}(pp\to \{pp\}_{\!s})
=N_{pp}\,4m^2<\psi_\mathbf{k}^{(-)}|V(^1S_0)|\mathbf{q}>  =}\\
&&-4m^2N_{pp}(\fmn{1}{2}\nu_1\fmn{1}{2}\nu_2|00) 4\pi\int_0^\infty
j_0(qr)V_{\!s}(r)\,\psi_\mathbf{k}^{(-)}(\mathbf{r})\,r^2\dd r,
\nonumber\\ \label{halfoff}
\end{eqnarray}
where $\nu_1$ and $\nu_2$ are the projections of the initial
proton spins. In Eq.~(\ref{halfoff}),
$\psi_\mathbf{k}^{(-)}(\mathbf{r})$ is the $pp$ scattering wave
function that is the solution of the Schr\"odinger equation with
the interaction potential $V(^1S_0)$ for a c.m.\ momentum ${|\bf
k|}$. It satisfies the following asymptotic boundary condition:
\begin{equation}
\label{boundary} \psi_\mathbf{k}^{(-)}(\mathbf{r})\to
\frac{\sin(kr+\delta)}{kr}\,,
\end{equation}
where $\delta$ is the $^{1\!}S_0$ phase shift. For simplicity of
presentation, we omit here the Coulomb interaction, though this is
taken into account in the actual numerical calculations. The
combinatorial factor $N_{pp}=2$ takes into account the identity of
the two protons.

The amplitude for the triangle diagram in Fig.~\ref{figope1}d is
given by the following four--dimensional integral

\begin{widetext}
\begin{eqnarray}\label{triangle}
A^{triangle}(pd\to \{pp\}_{\!s}n)&=& \int \frac{\dd^3p_1 \dd
T_1}{i(2\pi)^4} \sum_{\nu_1 \nu_2}\frac{A_{\nu_p}^{\nu_1}(p\to
\pi^0 p_1)
 A_{\lambda}^{\nu_2 \nu_n}
(\pi^0d\to pn)A_{\nu_1 \nu_2}(pp\to \{pp\}_{\!s})} {(2m)^2
(m_{\pi}^2-k_{\pi}^2-i\varepsilon)
 (\mathbf{p}_1^2/2m-T_1-i\varepsilon)  (\mathbf{p}_2^2/2m-T_2-i\varepsilon)},
\end{eqnarray}
\end{widetext}
where $T_i,\, \mathbf{p}_i, \,\nu_i$ are the kinetic energy,
three--momentum and projection of the spin of the intermediate
\emph{i}'th proton ($i=1,2$), respectively. Closing the contour of
integration in the lower--half $T_1$ plane, and taking into
account the residue at the point $T_1=\mathbf{p}_1^2/2m
-i\varepsilon $, one finds from Eq.~(\ref{triangle}) that
\begin{eqnarray}\nonumber
\lefteqn{A^{triangle}(pd\to\{pp\}_{\!s}n)=
-N_{pp}\sum_{\nu_1\nu_2} (\fmn{1}{2}\nu_1\fmn{1}{2}\nu_2|00)}
\\ \nonumber
&&\times \int \frac {\dd^3 p_1}{(2\pi)^3} \frac{A_{\lambda}^{\nu_2
\nu_n} (\pi^0d\to pn)<\psi_\mathbf{k}^{(-)}|V(^1S_0)|\mathbf{q}>m}
{ (m_{\pi}^2-k_{\pi}^2-i\varepsilon)
 (\mathbf{q}^2-\mathbf{k}^2-i\varepsilon) }\\
&&\times\, A_{\nu_p}^{\nu_1}(p\to \pi^0 p_1).\label{triangle2}
\end{eqnarray}

The pole diagram with an intermediate $\pi^0$ meson depicted in
Fig.~\ref{figope1}e leads to the following amplitude
\begin{equation}
\label{pole}
A^{pole}(pd\to\{pp\}_{\!s}n)=\frac{A_{\nu_p}^{\nu_{p_1}}(p\to \pi^0
p_1) A_{\lambda}^{\nu_{p_2}\nu_n} (\pi^0d\to
pn)}{m_{\pi}^2-k_{\pi}^2-i\varepsilon},
\end{equation}
where $\nu_{p_1}$ and $\nu_{p_2}$ are the projections of the spins
of the two final protons that are in the $^1S_0$ state. There is
another pole diagram with an intermediate $\pi^+$ meson but this
can be safely neglected here because it does not lead to low
energy $pp$ pairs.

Making the coherent sum of the triangle and the properly
antisymmetrized pole  amplitudes, given respectively by
Eq.~(\ref{triangle2}) and Eq.~(\ref{pole}), we find
\begin{widetext}
\begin{eqnarray}\nonumber
\lefteqn{A_{\nu_p
\lambda}^{\nu_n}(pd\to\{pp\}_{\!s}n)=A^{triangle}+A^{pole}=}
\\
&&N_{pp}\frac{f_{\pi NN}}{m_{\pi}}2m F_{\pi NN}(k_{\pi}^2)
 \sum_{\nu_1 \nu_2}
(\fmn{1}{2}\nu_1\fmn{1}{2}\nu_2|00)
A_{\lambda}^{\nu_{p_2}\nu_n}(\pi^0d\to pn) \int\frac{\dd^3
p_1}{(2\pi)^3}\frac{<\chi_{\nu_1} |(\bfg \sigma \cdot \mathbf{Q})
|\chi_{\nu_p}>\Psi^{(-)^*}_\mathbf{k}(\mathbf{q})}
{m_{\pi}^2-k_{\pi}^2-i\varepsilon}\:\cdot\label{tpluspol}
\end{eqnarray}
\end{widetext}
We have here used  the Lippmann--Schwinger equation
\begin{equation}
\label{psiq}
\psi_\mathbf{k}^{(-)^*}(\mathbf{q})=(2\pi)^3\delta^{(3)}(\mathbf{q}-\mathbf{k})
-\frac{m <\psi_\mathbf{k}^{(-)}|V(^1S_0)|\mathbf{q}>}
{\mathbf{q}^2-\mathbf{k}^2-i\varepsilon}\:\cdot
\end{equation}
The integral over ${\mathbf p}_1$ in Eq.~(\ref{tpluspol}) can be
evaluated in the rest frame of the final diproton, where ${\mathbf
p}_1={\mathbf q}$, as was done for the $pd\to dp$
reaction~\cite{kolybsm}. With this in mind, the kinematic
variables $\mathbf{Q}$ and the pion propagator are rewritten as:
\begin{eqnarray}
\nonumber \mathbf{Q}&=&\sqrt{\frac{E_p+m}{2m}}\left
\{(\mathbf{p}_1-{ \mathbf{p}_{II}})+ \mathbf{R}\right \},\\
\nonumber { \mathbf{p}}_{II} &=&\frac{\mathbf{p}_p}{E_p/m}, \ \
\mathbf{R}=-\frac{m}{E_p}\frac{T_p}{E_p+m}\mathbf{p}_p,\\
\label{Q} k^2-m_{\pi}^2&=&-\frac{E_p}{m}\left \{(\mathbf{p}_1-{
\mathbf{p}_{II}})^2+ \delta^2_{II}\right \},\ \\ \nonumber
\delta^2_{II}&=&
\frac{T_p^2}{(E_p/m)^2}+\frac{m_{\pi}^2}{E_p/m},
\end{eqnarray}
where $E_p$, $\mathbf{p}_p$ and $T_p=E_p-m$ are the total energy,
three--momentum and kinetic energy of the initial proton in the
rest frame of the final diproton.

Values of the $pd\to \{pp\}_{\!s}n$ cross section were presented
in Ref.~\cite{komarov2003} with a cut--off in the $pp$ excitation
energy of $E_{pp}^{max}=3\,$MeV. Defining the corresponding
maximum relative momentum through $k_{max}=\sqrt{m
\,E_{pp}^{max}}$, the c.m.\ differential cross section
becomes~\cite{smiruz}
\begin{eqnarray}\nonumber
\lefteqn{\frac{\dd\sigma}{\dd\Omega_n}(pd\to
\{pp\}_{\!s}n)=\frac{1}{(4\pi)^5}\frac{p_f}{p_i} \int
_0^{k_{max}}\dd k \frac{k^2}{s_{pd} \sqrt{m^2+\mathbf{k}^2}}\,}\\
&&\times\,\frac{1}{2}\int \dd\Omega_\mathbf{k}\,{\overline
{|A(pd\to \{pp\}_{\!s}n)|^2}}.\phantom{xxxxxxxxxxxxxxxxxxx}
\label{2to3}
\end{eqnarray}
The factor of $1/2$ in front of the angular integration in
Eq.~(\ref{2to3}) takes into account the identity of two final protons.

We choose the reference frame where the final diproton is at rest
and let the quantization axis $OZ$ lie along the direction of the
initial proton $\mathbf{p}_p$. In this frame only the longitudinal
components ($\mu=0$) of the vectors $\mathbf{p}_{II}$ and
$\mathbf{R}$ are non--zero. Thus the spin--averaged--squared
amplitude of the $pd\to \{pp\}_{\!s}n$ reaction can be written in
the following factorized form:
\begin{eqnarray}
\nonumber \lefteqn{{\overline
{|A^{triangle}+A^{pole}|^2}}=\frac{1}{4}|N_{pp} \frac{f_{\pi
NN}}{m_{\pi}}2m\,F_{\pi NN}(k_{\pi}^2)|^2} \nonumber
\\ \nonumber
&&\times \Bigl |\int \frac{\dd^3{q}}{(2\pi)^3}
\frac{Q^{\mu=0}}{k^2_{\pi}-m_{\pi}^2+i\varepsilon}\,
\psi_{k}^{(-)^*}(\mathbf{q})\Bigl |^2\: {\overline {|A(\pi^0 d\to
pn)|^2}},\\\label{ampsquared}
\end{eqnarray}
where ${\overline {|A(\pi^0 d\to pn)|^2}}$ is the
spin--averaged--squared amplitude of the $\pi^0d\to pn$ reaction.
This factorization is a consequence of the simple spin structure
of the diproton vertex $pp\to \{pp\}_{\!s}$.

For the $^{1\!}S_0$ final state,
${\overline{|A(pd\to\{pp\}_{\!s}n)|^2}}$ does not depend upon the
direction of the proton momentum $\mathbf{k}$ in the diproton rest
frame, so that the integration over $\dd\Omega_\mathbf{k}$ merely
gives a $4\pi$ factor. The cross section can be finally written
as:
\begin{eqnarray}
\nonumber \lefteqn {\frac{\dd\sigma}{\dd\Omega_n}^{OPE-II}(pd\to
\{pp\}_{\!s}n) =}\\ \nonumber &&\frac{1}{3\pi^2}\frac{p_f}{p_i}
\frac{q_{pn}}{q_{\pi d}} \frac{s_{pn}}{s_{pd}} \Bigl [
\frac{f_{\pi NN}}{m_{\pi}}2{m} F_{\pi NN}(k_{\pi}^2)
  \Bigr ]^2\,\frac{\dd\sigma}{\dd\Omega}(pn\to d\pi^0)
\\
&&\label{difpdppn} \times
 \int_0^{k_{max}} \dd k \frac{k^2}{\sqrt{m^2+k^2}}\,
|J^{\mu=0}_{pp}({ p_{II}}, \delta_{II})|^2.
\end{eqnarray}

The form factor $\mathbf{J}_{pp}$ is defined through
\begin{eqnarray}
\nonumber \lefteqn{\mathbf{J}_{pp}(p_{II},\delta_{II})=\int
\frac{\dd^3{q}}{(2\pi)^3}
\frac{\mathbf{Q}}{k^2_{\pi}-m_{\pi}^2+i\varepsilon}\,
\psi_{k}^{(-)^*}(\mathbf{q})}
\\
&&=\sqrt{\frac{E_p+m}{2m}} \frac{m}{E_p}\left
\{\mathbf{R}F_0(p_{II},\delta_{II})-i{\hat {\bf p}_p}\,
\Phi_{k}^{pp}(p_{II},\delta_{II})\right\},\nonumber\\
\label{ffdiproton}
\end{eqnarray}
\vspace{-0.3cm}
\begin{eqnarray}
\label{f0}
{\hspace{-1cm}F_0=\int_0^\infty \dd r\,r\,j_0(p_{II}\,r)\exp{(-\delta_{II}
r)}\,\psi_\mathbf{k}^{(-)^*}(\mathbf{r}),}
\end{eqnarray}
\vspace{-0.3cm}
\begin{eqnarray}
\Phi_{k}^{pp}=i\int_0^\infty \dd r (\delta_{II} r+1)
j_1({p_{II}}r) \exp\{-\delta_{II} r\}\,
\psi_\mathbf{k}^{(-)^*}(\mathbf{r}),\nonumber\\ \label{phi10pp}
\end{eqnarray}
where the kinematic variables $\mathbf{R}, \, \mathbf{p}_{II},
\,\delta_{II}$ are determined by Eq.~(\ref{Q}).

\subsection{ The $pd\to dp$ reaction}
\label{32} The OPE-II diagram for the reaction $pd\to dp$,
depicted in Fig.~\ref{figope1}c, includes two contributions
corresponding to a $\pi^+$ ($A^+$) and a $\pi^0$ ($A^0$) in the
intermediate state. Using isospin invariance, the coherent sum of
these diagrams is equivalent to that with $\pi^0$ multiplied by an
isospin factor of 3: $A^+ + A^0=3A^0$.

For the $ pn\to d$ vertex one has
\begin{equation}
\label{dpn} A^{\lambda'}_{\nu_p\nu_n}(pn\to d)=-4m\sqrt{m}
\left(\varepsilon+\frac{\mathbf{q}^2}{m}\right)
\varphi_{\lambda'}^{{\nu_p\nu_n}^*} (\mathbf{q}),
\end{equation}
where  $\varphi_{\lambda'}^{\nu_p\nu_n} (\mathbf{q})$ is the
deuteron wave function in momentum space
\begin{eqnarray}
\varphi_{\lambda'}^{\nu_p\nu_n} (\mathbf{q})=
\sum_{L,M_L,M_S}(\fmn{1}{2}\nu_p \fmn{1}{2}\nu_n|1M_S) \nonumber
  \\ \label{dwf}
\times (LM_L1M_S|1\lambda)\,Y_{LM_L}({\hat{\bf q}})\,u_L(q),
\end{eqnarray}
with  Clebsch--Gordan coefficients and spherical harmonics in
standard notation and
  $u_0(q)$ and $u_2(q)$ being respectively the $S$-- and
$D$--state components. The wave function is normalized as
\begin{eqnarray}
\frac{1}{3}\sum_{\nu_p\,\nu_n\, \lambda}\int
\frac{\dd^3q}{(2\pi)^3} | \varphi_{\lambda}^{\nu_p\nu_n}
(\mathbf{q})|^2 \nonumber \\ \label{normadwf}
=\int_0^\infty\Bigl [u_0^2(q)+u_2^2(q)\Bigr
]q^2\frac{\dd q}{(2\pi)^3}=1.
\end{eqnarray}

The total $pd\to dp$ transition amplitude becomes
\begin{eqnarray}
  \nonumber
\lefteqn{A_{\nu_p \lambda}^{\nu'_p \lambda'}(pd\to dp)=
3\frac{f_{\pi NN}}{m_{\pi}}F_{\pi NN}(k_{\pi}^2)
2\sqrt{m}}\\
&&\times\sum_{\nu_1,\nu_2,\mu} \sqrt{3}(1\mu
\fmn{1}{2}\nu_p|\fmn{1}{2}\nu_1)
 \int \frac{\dd^3q}{(2\pi)^3}
\frac{{\tilde Q}^{\mu}}{k^2-m_{\pi}^2+i\varepsilon}\nonumber \\
&&\times\, \varphi_{\lambda'}^{\nu_1\nu_2}
(\mathbf{q})A_{\lambda}^{\nu_2\nu_p'} (\pi^0 d\to pn),
\label{mpdelas1}\end{eqnarray}
where $\nu_p$ ($\nu_p'$)
and$\lambda$ ($\lambda'$) are the spin projections of the initial
(final) proton and deuteron.

The integral over the three--momentum ${\mathbf q}$
of the intermediate nucleon
is evaluated in the rest frame of the final deuteron.
There it takes the form
\begin{eqnarray}
\nonumber J_L^{\mu}({\tilde p}, \delta)= \int
\frac{\dd^3q}{(2\pi)^3} \frac{{\tilde Q}^{\mu}\,
\varphi_{\lambda}^{\nu_1\nu_2}(\mathbf{q})}
{k^2-m_{\pi}^2+i\varepsilon} =\nonumber
\\
\sqrt{\frac{E_p+m}{2m}} \frac{m}{E_p}  \label{intd} \left\{
R^{\mu}F_L({\tilde p}, \delta)-i{\hat{\bf p}}_p^{\mu}
\Phi_{1L}({\tilde p},{\tilde \delta})\right \},
\end{eqnarray}
where the quantization axis is chosen to lie along $\mathbf{p}_p$.
The kinematical variables $\tilde{\mathbf Q}$, ${\tilde p}$ and
${\tilde \delta}$ come from Eqs.~({\ref{Q}) for the variables
${\mathbf Q}$, $p_{II}$ and $\delta_{II}$, with $E_p$,
$\mathbf{p}_p$ and $T_p$ being replaced respectively by the total
energy  ${\tilde E}_p=\sqrt{m^2+{\mathbf{\tilde{ p}}_p^2}}$,
three--momentum $\mathbf{\tilde {p}}_p$, and kinetic energy
$\tilde {T}_p=\tilde{ E}_p-m$ of the initial proton in the rest
frame of the final deuteron. The form factors $F_L$ and
$\Phi_{lL}$ are defined by Eqs.~(\ref{phi02}).

Finally, the c.m.\ $pd\to dp$ differential cross section is
predicted to be:
\begin{eqnarray}
\frac{\dd\sigma}{\dd\Omega}^{OPE-II} (pd\to dp)=9 \Bigl [
\frac{f_{\pi NN}}{m_{\pi}}2\sqrt{m} F_{\pi NN}(k_{\pi}^2)
  \Bigr ]^2 \frac{s_{pn}}{s_{pd}}
\nonumber
\\ \times\frac{q_{pn}}{q_{\pi d}}
 \Bigl \{|J^{\mu=0}_0({\tilde p}, {\tilde
\delta})|^2+
 |J^{\mu=0}_2({\tilde p},{\tilde  \delta})|^2 \Bigr \}
\frac{\dd\sigma}{\dd\Omega}(pn\to d\pi^0).\nonumber \\\label{difpddp}
\end{eqnarray}
For backward proton--deuteron elastic scattering, the $pp\to
d\pi^+$ cross section is also to be taken for a similar
forward--going deuteron. Since Eq.~(\ref{difpddp}) coincides with
Eq.~(1) of Ref.~\cite{vegh}, the OPE-II and OPE-I models give the
same formula for the unpolarized $pd \to dp$ cross section, as
required.

On the basis of Eqs.~(\ref{difpdppn}) and (\ref{difpddp}), we can
find the following factor relating the $pd\to \{pp\}_{\!s}n$ and
$pd\to dp$ differential cross sections to be compared to that in
Eq.~(\ref{relation2}):
\begin{eqnarray}
\lefteqn{R_{II}=}\nonumber\\
\nonumber &&\hspace{-3mm} \left.\frac{m}{27\pi^2}
\int_0^{k_{max}}\dd k
\frac{k^2}{\sqrt{m^2+k^2}}|\Phi_{k}^{pp}(p_{II},\delta_{II})|^2\right/
|\Phi_{10}^{d}({\tilde p},{\tilde \delta}) |^{2},\\ \label{R2}
\end{eqnarray}
where the integrals $\Phi_{10}^{d}({\tilde p},{\tilde \delta})$
and $ \Phi_{k}^{pp}( p_{II},\delta_{II})$ are determined by
Eqs.~(\ref{phi02}) and (\ref{phi10pp}), respectively.
Approximating the integral in Eq.~(\ref{R2}) by using the value of
the integrand at ${\overline E_{pp}}=E_{pp}^{max}/2$, one can rewrite the
equation as
\begin{equation}
\label{R2a}
R_{II}\approx\frac{2}{27}\frac{k_{max}^3}{6\pi^2 m}
\frac{|\Phi_{\overline k}^{pp}(p_{II},\delta_{II})|^2}
{|\Phi_{10}^{d}({\tilde p},{\tilde \delta})|^2},
\end{equation}
where $\overline k=\sqrt{2\,m\,\overline E_{pp}}$. In the
derivation of $R_{II}$ we have neglected the contribution of the
deuteron $D$--state component and the form factor $F_L$, which are,
however, included in the numerical evaluations.

The origins of the different terms in Eq.~(\ref{R2a}) are easy to
understand. To obtain Eq.~(\ref{difpddp}) from (\ref{difpdppn})
one needs to make the following replacements: (i) $\psi_{\bf
k}^{(-)}({\bf r})\to \varphi_d({\bf r})/\sqrt{m}$; (ii) multiply
by the ratio of the isospin and combinatorial factors
$9/\left(N^2_{pp}/2\right)= 9/2$; (iii) multiply by the spin
factor of three; (iv) multiply by the factor ${4\pi^2}$, which
arises from the difference between three-- and two--body phase
spaces; (v) divide by the  factor
\begin{equation}
\label{sbeta} \int _0^{k_{max}}\frac{k^2}{\sqrt{m^2+k^2}}\dd k
\approx \frac{k_{max}^3}{3m}.
\end{equation}

\section{The exchange  of baryons with
$\mathbf{T=\frac{1}{2}}$ in the $\mathbf{t}$--channel}
\setcounter{equation}{0}

The baryon exchange (BE) amplitude for $pd\to\{pp\}_{\!s}n$ of
Fig.~\ref{fig3}c can be written as:
\begin{eqnarray}
\lefteqn{A^{BE}(pd\to\{pp\}_{\!s} n)= } \nonumber\\\label{BE}
&&\hspace{-5mm}\sum_{\nu_N}\frac {A_{\lambda}^{\nu_N\nu_n}(d\to
pN^*) A_{\nu_p\nu_N}(pN^*\to\{pp\}_{\!s}) }
{m_{N^*}^2-t-i\varepsilon},
\end{eqnarray}
where $m_{N^*}$ is the mass of the exchanged baryon, $\nu_N$ its
spin projection, $t=(p_d-p_n)^2$ the four--momentum transfer, and
$A_{\nu_p\nu_N}(pN^*\to\{pp\}_{\!s})$ and
$A_{\lambda}^{\nu_N\nu_n}(d\to pN^*)$ the amplitudes of the sub
processes $pN^*\to \{pp\}_{\!s}$ and the vertex $d\to pN^*$
respectively. While the case of one--nucleon--exchange can be
found in Refs.~\cite{smiruz,uzjpg,haidenuz2003}, the formalism for
$N^*$ with higher spins was studied in Ref.~\cite{sharmamitra74},
where a good fit to the cross section data on the $pd\to dp$ and
$pp\to d\pi^+$ reactions was obtained for beam energies
$T_p>1$\,GeV.

\begin{figure}[hbt]
\mbox{\epsfig{figure=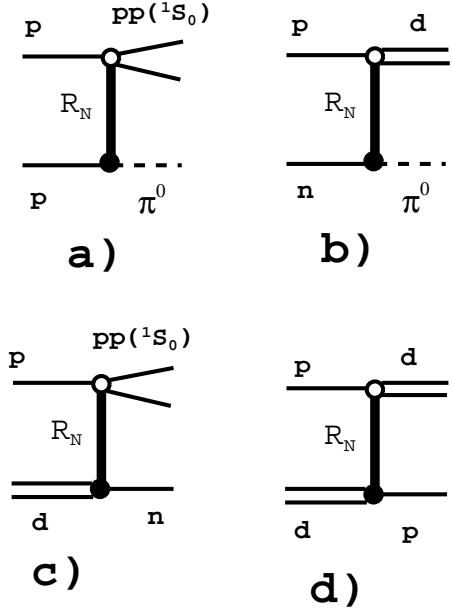,height=0.35\textheight, clip=}}
\caption{The exchange of baryons with isospin $T=\frac{1}{2}$ (N,
N$^*$, and Reggeon) in the $t$--channel of the
$pp\to\{pp\}_{\!s}\pi^0$, $pn\to d\pi^0$, $pd\to \{pp\}_{\!s}n$
and $pd\to dp$ reactions.} \label{fig3}
\end{figure}

For our present purposes, the main features of the BE mechanism
are (i) its isospin structure with $T=\frac{1}{2}$ in the
$t$--channel, and (ii) the factorized residue of the amplitude.
The same features are present in the Reggeon mechanism, where the
transition amplitude is given by
\begin{equation}
\label{Reggeon} A(s,t)=F(t)\Bigl(\frac{s}{s_{0}}\Bigr
)^{\alpha_N(t)} \exp { \Bigl [-i\frac{\pi}{2}
\left(\alpha_N(t)-\fmn{1}{2}\right)\Bigr ]},
\end{equation}
where $\alpha_N(t)$ is the nucleon Regge trajectory. The residues
of the Regge amplitudes $F(t)$ can be factorized into products of
terms coming from the upper and lower vertices of Fig.~\ref{fig3}.
Therefore, within the baryon or Reggeon exchange (BRE) model, one
obtains the following relation between the c.m.\ cross sections:
\begin{eqnarray}
\lefteqn{\frac{\dd\sigma}{\dd\Omega}^{BRE}(pd\to \{pp\}_{\!s}n)=}
\nonumber \\
&&\hspace{-5mm}\frac {\frac{\dd\sigma}{\dd \Omega}(pp\to
\{pp\}_{\!s}\pi^0)} {\frac{\dd\sigma}{\dd\Omega}(pn\to
d\pi^0)}\times \frac{\dd\sigma} {\dd\Omega}^{BRE}(pd\to
dp\label{reggerel}).
\end{eqnarray}
Here the  cross  sections, within the BRE model of
Fig.~\ref{fig3}, are taken at the same four--momentum transfer $t$
for all reactions and at $s=s_{pp}\approx s_{pn}$ for the $pn\to
d\pi^0$ and $pp\to \{pp\}_{\!s}\pi^0$, and $s=s_{pd}$ for the
$pd\to \{pp\}_{\!s}n$ and $pd\to dp$ reactions. In deriving this
relation we assume that the $t$-dependence of the vertices is
smooth. Formally Eq.~(\ref{reggerel}) coincides with
Eq.~(\ref{relation2}) with $R_I=1$.

\section{Results and Discussion }
\setcounter{equation}{0}

\begin{figure}[hbt]
\mbox{\epsfig{figure=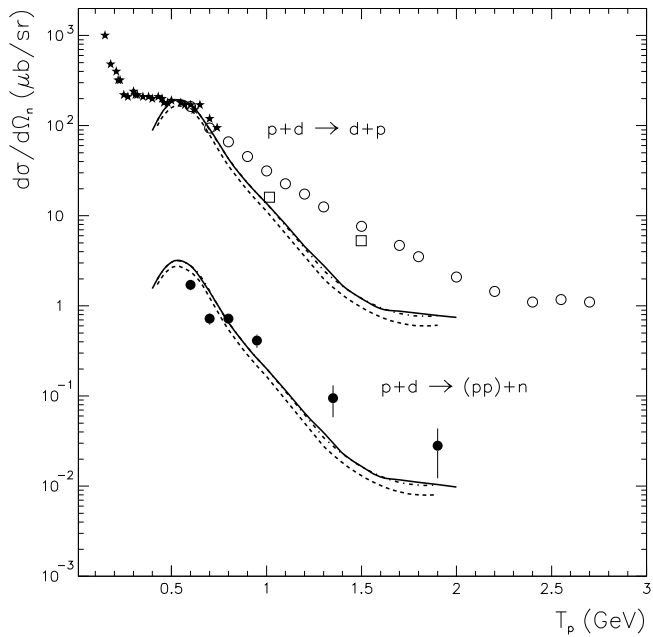,height=0.40\textheight, clip=}}
\caption {The differential cross sections for $pd \to dp$ at
$\theta_{c.m.}=180^o$ and $pd\to \{pp\}_{\!s}n$ averaged over
$\theta_{c.m.}=166^{\circ} - 180^{\circ}$ \emph{versus} the proton
beam energy compared with the predictions of the OPE-II model for
different values of the cut-off parameter: $\Lambda=1$\,GeV/c
(full line), 0.8 (dashed-dotted), 0.65\,GeV/c (dashed). The cross
section of the $pn\to d\pi^0$ reaction is taken from the SAID SP96
solution~\cite{SAID}. Data for $pd\to \{pp\}_{\!s}n$ and $pd\to
dp$ are those of Ref.~\cite{komarov2003} and
\cite{dubal,boudard,berthet} respectively.} \label{fig4}
\end{figure}

\subsection{The OPE-II model}
The results of our calculations are shown in Figs.~\ref{fig4} and
\ref{fig5}. For the $pd\to dp$ differential cross section, the
OPE-I and OPE-II approaches give identical results and they
reproduce the observed shoulder in the energy dependence in the
$T_p=0.5-0.7$\,GeV region, which is caused by virtual $\Delta$
excitation~\cite{cwilkin69,lak,boudardillig,uzikovrev,haidenuz2003}.
At higher energies, $T_p> 1$\,GeV, the OPE cross section falls
faster than the data. The calculated  cross sections varies very
weakly with increasing cut-off parameter $\Lambda$ in the $\pi NN$
vertex.

The OPE-II model for $pd\to \{pp\}_{\!s}n$ is in reasonable
agreement with the experimental data below 1\,GeV, being best at
about 0.8\,GeV. It is interesting to note that at this energy and
$\theta_{cm}=180^\circ$ the ONE mechanism vanishes due to a
repulsive core in the $NN$-interaction, as illustrated in
Fig.~\ref{fig5}~\cite{haidenuz2003}. As a result, double
scattering with the excitation of the $\Delta$(1232)-isobar was
found to be dominant in this region. Since $pn\to d\pi^0$ is also
$\Delta$--dominated in this region, the agreement between the
OPE-II model and the $pd\to \{pp\}_{\!s} n$ data seems largely to
confirm the results of Ref.~\cite{haidenuz2003}. Furthermore, at
this kinematic point the ONE amplitude changes sign, as does the
ONE--OPE interference.

Outside this region, the ONE mechanism gives a sizable
contribution~\cite{uzjpg,haidenuz2003}, which suggests that the
disagreement between the data and the OPE-II  model away from
$T_p\approx0.8$\,GeV may be connected with the ONE contribution.
In Fig.~\ref{fig5} we show the ONE (DWBA) contribution taken from
Ref.~\cite{haidenuz2003} and its coherent sum with the OPE
contribution, with the relative sign being chosen to get the best
agreement with the data~\cite{komarov2003}. We are here implicitly
assuming that ONE is negligible in the physical $\pi^0d\to pn$
amplitude.

Above 1\,GeV, the cross section for the $pd\to \{pp\}_{\!s}n$
reaction calculated in the OPE-II model falls faster than the data
with increasing energy. In this model the energy slope for both
this and the $pd\to dp$ reaction is determined mainly by the
energy dependence of the cross section of the $pn\to d\pi^0$
reaction; other kinematic factors and form factors are very smooth
functions of the beam energy. As a result, the ratio of diproton
to deuteron formation is practically independent of $T_p$.

As explained in Sec. \ref{32}, the strong preference for deuteron
formation within the OPE-II mechanism is the result of several
considerations, including spin--isospin, combinatorial, phase
space factors as well as the ratio of form factors in
Eq.~(\ref{R2}). For a maximum diproton excitation energy of
$E_{pp}^{max}=3$\,MeV and beam energy in the interval $0.6
-1.9$\,GeV, Eqs.~(\ref{R2},\,\ref{R2a}) predict a ratio of
$R_{II}\approx 0.016 - 0.013$, which is in qualitative agreement
with the experimental value $R^{exp}=0.010 -
0.011$~\cite{komarov2003}.

In contrast to the OPE-II model, within the OPE-I formalism of
Eq.~(\ref{ratio1}) the small magnitude of the ratio $R_I$ follows
mainly from the small ratio of the cross sections of the $pp\to
\{pp\}_{\!s}\pi^0$ and $pn\to d\pi^0$ reactions, as seen from
Ref.~\cite{dymov06} at 0.8\,GeV. Results within this approach will
remain ambiguous until there is more information on the $pn\to
\{pp\}_{\!s}\pi^-$ amplitude.
\begin{figure}[hbt]
\mbox{\epsfig{figure=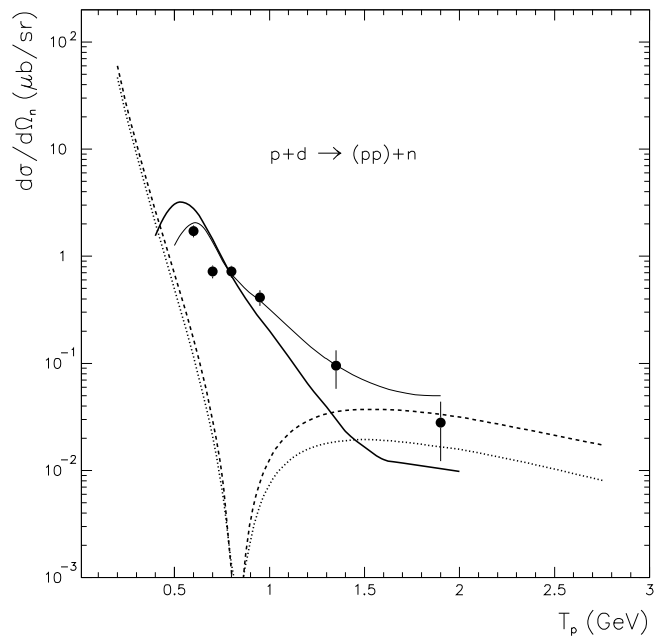,height=0.40\textheight, clip=}}
\caption{Differential cross section for the $pd\to \{pp\}_{\!s}n$
reaction~\cite{komarov2003}. The full thick curve shows the OPE-II
results for $\Lambda=1$\,GeV/c. The
predictions~\cite{haidenuz2003} of the ONE(DWBA) mechanism with
the CD Bonn potential are shown by the dashed (Born approximation)
and dotted (with distortions) curves. The coherent sum of the
OPE-II  and the ONE(DWBA) is shown by the thin full line. }
\label{fig5}
\end{figure}
\subsection{The OPE-I  and BRE models}
At present the OPE-I approach can only be compared with the
$pd\to\{pp\}_{\!s}n$ data at 0.8\,GeV, where results on the
$pp\to\{pp\}_{\!s}\pi^0$ have recently appeared~\cite{dymov06}.
Assuming that the BRE mechanism of Fig.~\ref{fig3mech}c dominates
the $pN\to \{pp\}_{\!s}\pi$ amplitude at this energy, we find from
Eq.~(\ref{relation2}) a value of the $pd\to \{pp\}_{\!s}n$
differential cross section of $0.7 \mu\textrm{b}/sr$, which is in
a good agreement with the data~\cite{komarov2003}. On the other
hand, if the $\Delta-$isobar mechanism dominates pion production
at 0.8\,GeV~\cite{niskanen06}, then the OPE-I approach falls too
low by a factor two. Graphs with an intermediate $N^*$, as in
Fig.~\ref{fig3mech}b, would make the underestimate a factor of
nine.

If the BRE mechanism is indeed important for the $pp\to
\{pp\}_{\!s}\pi^0$ reaction at 0.8\,GeV, one should analyze the
role of this mechanism also in the $pn\to d\pi^0$, $pd\to
\{pp\}_{\!s}n$ and $pd\to dp$ reactions. Using the $pp\to
\{pp\}_{\!s}\pi^0$ data~\cite{dymov06} and the SAID SP96
solution~\cite{SAID} for the $pn\to d\pi^0$ reaction, we find from
Eq.~(\ref{reggerel}) that the BRE model also predicts the same
value of $0.7\,\mu$b/sr for the $pd\to \{pp\}_{\!s}n$ cross
section. Within the Reggeon model, the small magnitude of the
$pd\to \{pp\}_{\!s}n$ cross section, as compared to the $pd\to
dp$, should be considered to be consequence of the relative sizes
of the residue functions at the $pR_N\{pp\}$ and $pR_N d$
vertices.

In order to get more insight into the dynamics of the $pd\to
\{pp\}_{\!s}n$ and $pp\to \{pp\}_{\!s}\pi^0$ reactions one has to
discriminate between the BRE and the $\Delta$-isobar mechanism of
the reaction $pp\to \{pp\}_{\!s}\pi^0$ at 0.8\,GeV (and higher
energies). For this purpose it is important to measure the
unpolarized cross section of the $pn\to \{pp\}_{\!s}\pi^-$
reaction since
\begin{equation}
\label{deltaNNpi} \frac{\dd\sigma}{\dd\Omega}(\pi^0)/\frac{\dd
\sigma}{\dd\Omega}(\pi^-)=
\begin{cases} 2, \ {\text{ $\Delta-$mechanism,}} \\
 \fmn{1}{2},
 \ {\text { $T=\fmn{1}{2}$ $t$--channel exchange. }}\\
\end{cases}
\end{equation}

\subsection{ The Reggeon mechanism and
constituent--quark counting rules}

We have shown that the OPE-II model can explain the similarity in
the energy dependence of the $pd\to dp$ and $pd\to \{pp\}_{\!s}n$
cross sections but underestimates both of their overall values at
$T_p=1-2$\,GeV.  It was argued that this discrepancy might be due
to contributions from ONE or baryon (Reggeon) exchanges. If this
is true, it would mean that the effective degrees of freedom in
these reactions are non--nucleonic. In this connection it is
interesting to check whether the constituent-quark counting rules
(CCR)~\cite{MMT,BF} can be applied to these reactions. A scaling
behavior related to the CCR was observed in the $\gamma d\to pn$
reaction at photon beam energy  1--4~GeV (see
Refs.~\cite{rossi05,grishina} and references therein). Recently
the CCR  behavior was found also in the  $pd\to dp$ and $dd\to
\,^{3}\textrm{H}p$ reactions in the GeV energy region at large
scattering angles~\cite{uzccr}. This suggests that one might
usefully search for a similar CCR behavior in the $pd\to dp$ and
$pd\to\{pp\}_{\!s}n$ reactions, at least in the region between the
$\Delta(1232)$ and $\Delta (1920)$ resonances, say between 1 and
2\,GeV.

According to the CCR hypothesis, the energy dependence of the
invariant cross sections can be parameterized as
\begin{eqnarray}
\label{ccr} \frac{\dd\sigma}{\dd t}=\frac{\pi}{p_i p_f}
\frac{\dd\sigma\phantom{w}}{\dd\Omega_{cm}} =
\frac{1}{s^n}f({\theta_{cm}}),
\end{eqnarray}
where the function $f({\theta_{cm}})$ does not depend on energy
and $n+2$ is the sum of all active point-like constituents in the
initial and final states. Our fit to the data shown in
Fig.~\ref{fig6} gives $n=12.9$ for both the $pd\to \{pp\}_sn$ and
$pd\to dp$ reactions, whereas CCR would suggest that
$n=3+6+3+6-2=16$. One would therefore require significant diquark
configurations in order to get better numerical agreement.

\begin{figure}[hbt]
\mbox{\epsfig{figure=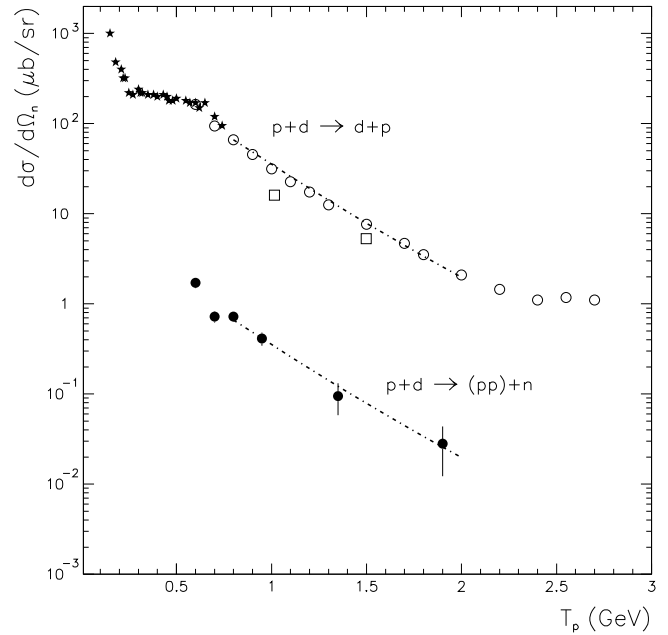,height=0.40\textheight, clip=}}
\caption{Differential cross sections for the $pd\to dp$ and $pd\to
\{pp\}_{\!s}n$ reactions as shown in Fig.~\ref{fig4}. The
dashed--dotted lines give the results of fitting the data within
the CCR approach of Eq.~(\ref{ccr}), where the invariant cross
section behaves as $\frac{\dd\sigma}{\dd t}=\emph{const}\times
s^{-12.9}.$ } \label{fig6}
\end{figure}

\section{Conclusions}

The present analysis shows that there are close connections
between the different reactions which lead to diproton formation
in the final state in $pd$ and $pN$-collisions. However, the
actual relations depend on the reactions mechanisms. We found that
the predictions of the OPE-II model, which is based on the $\pi^0
d \to pn$ sub process, are quite close to the $pp\to
\{pp\}_{\!s}n$ deuteron breakup data. This model allows us to
explain the absolute value of the $pd\to \{pp\}_{\!s}n$ cross
section at $\theta_{cm} \approx 180^{\circ}$ in the
$\Delta$--isobar region 0.6 -- 0.9\,GeV as well as its energy
dependence. It also describes the small value of the ratio
$R=\dd\sigma(pd\to \{pp\}_{\!s}n)/\dd\sigma (pd \to dp)$ in the
whole interval 0.6 -- 1.9\,GeV of measurement reported in
Ref.~\cite{komarov2003}.

The agreement points to an important contribution coming from the
$\Delta-$isobar below 1\,GeV, which enters \emph{via} the
$\pi^0d\to pn$ sub process but, on the other hand, suggests that
the ONE mechanism is relatively unimportant. To a large extent,
these conclusions are compatible with the results of the previous
analysis of this reaction, performed on the basis of a different
model~\cite{haidenuz2003}. The minor role found for the ONE
contribution sheds some light on the $T_{20}$ puzzle, discussed in
the introduction, which is entirely based on the assumption that
the ONE mechanism dominates the large momentum--transfer $pd$
reactions.

There is as yet insufficient information to describe the
$pd\to\{pp\}_{\!s}n$ data unambiguously within the OPE-I model.
However, if we assume the dominance of $T=\fmn{1}{2}$ exchange in
the $pN\to \{pp\}_{\!s}\pi$ amplitude, as given for example by
baryon or Reggeon exchange, then a satisfactory description can be
achieved. Much of this ambiguity will be removed once data are
available from the forthcoming  measurements of the cross sections
for $pp\to \{pp\}_{\!s}\pi^0$ and $pp\to \{pp\}_{\!s}\pi^-$ at
$\theta_{cm}\approx 0^\circ$ in the 1 -- 2\,GeV
region~\cite{kulikov}.

\section*{Acknowledgments} The authors are grateful to
J.~Niskanen for useful discussions.  Two of the authors (YuU and
CW) wish to recognize the hospitality of the Institut f\"ur
Kernphysik of the Forschungszentrum J\"ulich, where much of this
work was carried out. This work was supported in part by the
Heisenberg--Landau programme.

\end{document}